# Detecting stimulated-Raman responses of few molecules using Optical Forces

Venkata Ananth Tamma[1], Lindsey M. Beecher[2], Jennifer S. Shumaker-Parry[2] and H. Kumar Wickramasinghe[3]

[1]Department of Chemistry, University of California, Irvine, CA 92697, USA
[2] Department of Chemistry, University of Utah, 315 S. 1400 E. Rm. 2020, Salt Lake City, UT 84112, USA
[3] Department of Electrical Engineering and Computer Science, University of California, 142 Engineering Tower, Irvine, CA 92697, USA
*Corresponding author: hkwick@uci.edu

*Abstract*: We demonstrate the stimulated Raman near-field microscopy of few molecules, measured only using near-field optical forces thereby eliminating the need for far-field optical detection. The molecules were excited in the near-field without resonant electronic enhancement. We imaged gold nanoparticles of 30 nm diameter functionalized with a self-assembled monolayer (SAM) of 4-nitrobenzenethiol (4-NBT) molecules. The maximum number of molecules detected by the gold-coated nano-probe at the position of maximum field enhancement could be fewer than 20 molecules. The molecules were imaged by vibrating an Atomic Force Microscope (AFM) cantilever on its second flexural eigenmode enabling the tip to be controlled much closer to the sample thereby improving the detected signal-to-noise ratio when compared to vibrating the cantilever on its first flexural eigenmode. We also demonstrate the implementation of stimulated Raman nanoscopy measured using photon-induced force with non-collinear pump and stimulating beams which could have applications in polarization dependent Raman nanoscopy and spectroscopy and pump-probe nano-spectroscopy particularly involving infrared beam/s.

Tip Enhanced Raman Spectroscopy (TERS) has proven to be an important tool for nanoscale imaging with chemical specificity [1]-[5]. The Raman signals in TERS are due to the high field enhancement afforded using a sharp metal probe brought very close to the molecule/s under study. The spectral content of far-field scattered photons due to the near-field interaction between a sharp metal probe and nearby molecule/s are recorded using a spectrometer with a highly sensitive detector array. In contrast to TERS where the molecule/s under study are excited in the near-field, but the scattered photons are detected in the far-field by using the sharp metal probe as an optical antenna, Photon-Induced Force Microscopy (PIFM) is an all near-field measurement. Previously, we investigated a scheme for both near-field excitation and near-field detection of Raman vibrational responses without the need for far-field collection of photons [6], [7]. The detection of the Raman responses was mediated by optical forces, generated purely by a near-field interaction between a sharp metal probe and the molecule/s under study. The generated optical force was measured using an Atomic Force Microscope (AFM). Previously, PIFM was used to detect and image linear molecular resonances at the nanometer level [8], [9], [13], perform non-linear imaging and spectroscopy at the nanoscale [6], [7], [9]-[11] and to image propagating surface plasmon polaritons (SPPs) on a gold surface [12]. Furthermore, PIFM can image molecular resonances over a large range of wavelengths spanning from the visible to mid-IR wavelength regimes [13].

In these measurements, PIFM uses an AFM to detect photon induced optical frequency interaction forces between two, photon-induced, dipoles: one in the sample under study and another formed at the tip of a metal coated AFM probe. An AFM feedback control loop brings the metal-coated tip very close to the sample leading to an enhanced electric field in the near-field region between the tip and sample. This locally enhanced electric field amplifies the photon induced interaction forces between the sample and the apex of the metal-coated tip which are physically positioned very close to each other. By modulating the optical excitation/s to generate a photon induced force interaction resonant with an eigenmode of the AFM cantilever, Raman vibrational information can be detected using the AFM. PIFM leads to optical background free molecular nanoscale imaging with chemical specificity because the weak Raman signals are detected only by mechanical means instead of detecting photons. Recently, it has been demonstrated that the PIFM could be used to detect lateral (parallel to sample surface) optical forces [14], [15]. The ability of PIFM to detect both longitudinal (along tip axis) and lateral optical forces, potentially simultaneously, could allow for detection of Raman vibrational response of molecules in both longitudinal and lateral directions. This could have applications in simultaneously estimating the different tensor components of the Raman vibrational responses, potentially allowing estimation of the orientation of the molecule/s. Recently, the near-field force interaction between stimulated Raman excited molecules and a sharp metal probe was measured using an AFM but without the need for resonant electronic enhancement gain [6]. The tip to sample spacing was controlled using the higher order flexural eigenmodes of the AFM cantilever enabling the tip to come very close to the sample. This results in improved detection sensitivity when compared with the previous work on Raman PIFM [6]. Stimulated Raman vibrational spectra of photo-switchable azobenzene thiol and the α-amino acid *l*-phenylalanine were measured and found to agree well with published theoretical and experimental results.

The results presented in [6], [7] however, were obtained by Raman-PIFM measurements performed over large aggregates of molecules deposited on glass substrates. Motivated by the potential for nanoscale chemical information about interfaces to impact research broadly in areas ranging from catalysis and photovoltaics to sensor development, we investigated nanoscale imaging and spectroscopy of a monolayer of molecules assembled on nanoparticles using Raman-PIFM. In this paper, 4-nitrobenzene thiol (4-NBT) molecules were adsorbed as a self-assembled monolayer (SAM) on gold nanoparticles so that the signal-to-noise in the detected near-field optical force could be increased due to the plasmon enhancement of the optical electric field between the gold substrate and a gold coated AFM probe. Using 30 nm diameter gold nanoparticles functionalized with 4-NBT, we clearly observe Raman-PIFM signals from few molecules. We chose to image the vibrational response of 4-NBT at 1343 cm$^{-1}$ associated with the symmetric stretch of the nitro group. The wavelengths of both the pump (604 nm HeNe laser) and the stimulating laser beams (657-664 nm laser diode) were chosen such that they were non-resonant with any of the molecular electronic absorption peaks of 4-NBT. The Raman vibrational responses of the 4-NBT molecule, excited by the stimulated Raman process, were measured by the PIFM technique with nanometer resolution using an AFM as a force detector.

Similar to our previous work [6], we again chose to vibrate the AFM cantilever using one of its higher flexural modes of vibration to enable operating the AFM very close to the sample surface thereby improving the measured Raman-PIFM signal [6]. By vibrating the cantilever on one of its higher order flexural eigenmodes, we reduce the amplitude of cantilever oscillations allowing for the gold coated tip apex to be brought much closer to the molecule/s under study. Thus, the apex of the gold coated AFM probe experiences the optical interaction force, which decays rapidly as $d^{-4}$ with increasing tip-sample distance $d$, for a large fraction of its oscillation cycle thereby increasing the detected optical force. In this work, a gold coated silicon cantilever was excited at its second flexural resonance with the topography also recorded at the second flexure resonance frequency while the Raman-PIFM response was simultaneously measured on the first flexural resonance. The AFM cantilever oscillation amplitude in our experiments was typically 2-3 nm and never exceeded 4 nm.

We also demonstrate Raman-PIFM based on a new optical excitation geometry with non-collinear pump and probe beams. In previous studies, we recorded Raman vibrational responses of molecules when both the pump and stimulating beam were collinear and focused on the sample using a microscope objective with high numerical aperture in a geometry we refer to henceforward as inverted transmission geometry [6]-[11]. In the current study, we investigate the application of the Raman-PIFM technique when the pump and stimulating beams are not collinear but instead illuminate the tip-sample from different directions, with a 120° angle between them. This novel illumination geometry could have applications to polarization dependent Raman studies and could allow for a flexible experimental setup for certain samples. Furthermore, it could be useful in PIFM pump-probe studies; particularly in experiments using infrared beams along with another infrared or a visible beam. This geometry can allow for PIFM pump-probe experiments without need for any dichroic beam-splitters.

It has been demonstrated that for stimulated coherent optical spectroscopy techniques such as Stimulated Raman Scattering, particularly for photon detection schemes, the measured material response is sensitive to the actual geometry of the experiment [16]. In particular, for experimental geometries involving tightly focused optical beams, the spatially varying phase of the excitation beams needs to considered [16]. However, in the particular case of the tip of the gold coated AFM probe interacting with a gold nanoparticle, we observe that the direction of the electric field is primarily dictated by the tip-nanoparticle gap. We therefore expect to measure the Raman vibrational responses of molecules when both the pump and stimulating beam are made non-collinear with each beam individually focused on the tip (the optical antenna) from different directions. In this work, we demonstrate Raman nanoscopy of few molecules when the pump beam is focused on the tip-sample junction using the inverted transmission geometry approach while the stimulating beam is focused on the tip from a completely different direction using a parabolic mirror.

**Experiment setup**

In Fig. 1 (a), we present the schematics of the experimental setup used to measure Raman-PIFM in the inverted transmission geometry. This PIFM geometry was used in previous investigations [6]-[11]. Briefly, two optical beams, a pump beam at frequency $v_p$ and a stimulating beam at frequency $v_s$, were made collinear and focused using an oil immersion objective (Olympus PlanApo 100x, with $NA$ = 1.45) to a diffraction limited spot on the sample surface. To ensure the tightest focal spot, the back aperture of the objective was over-filled. Both beams were linearly (along the x-axis) polarized and the focal spots of each beam were verified using PIFM as detailed in previous work [11] by recording the optical force measured between the gold coated AFM probe and its image dipole on a glass substrate. A radial polarizer could be used to further enhance the field along the z-axis between the tip apex and the sample, although in these experiments we chose not to do so due to adequate signal strength. Using a piezo mirror, the double lobes (spatial regions with purely longitudinal polarization along the z-axis [11]) of each focal spot were spatially overlapped. The apex of the gold coated tip was positioned over one of the double lobes, with purely longitudinal polarization along the z-axis, to create an enhancement of the electric fields along the z-axis between the tip apex and the sample.

The new non-collinear Raman-PIFM configuration with a combined inverted transmission and side illumination geometry is shown in Fig. 1 (b). A pump beam at frequency $v_p$ was focused using an oil immersion objective (Olympus PlanApo 100x, with $NA$ = 1.45) to a diffraction limited spot on the sample surface. To ensure the tightest focal spot, the back aperture of the objective was over-filled. The pump beam was linearly (along the x-axis) polarized and the focal spot of the beam was verified with PIFM using double lobes. The apex of the gold coated tip was positioned over one of the double lobes of the pump beam to create an enhancement of the electric fields due to the pump beams along the z-axis between the tip apex and the sample. A stimulating beam at frequency $v_s$, was then focused on the tip using an off-axis parabolic mirror with $NA = 0.5$ at an angle of 30° from the horizontal. The polarization of the stimulating beam was chosen to be parallel with the axis of the tip. The focusing of the stimulating beam from the side using a parabolic mirror was verified using the PIFM technique. We first approached the tip to the sample (on a glass substrate) and focused the light using the parabolic mirror onto the tip-sample junction. With the tip approaching the sample, the parabolic mirror was adjusted using piezo-electric actuators while the instantaneous PIFM signal observed on the lock-in read-out was maximized. In all experiments, the sample was raster scanned to obtain topographic information and spatially dependent stimulated Raman spectral information at the frequency $\Delta v = v_p - v_s$ simultaneously. On its second flexural resonance $f_{0,2}$, the cantilever was excited while the modulation frequencies $f_1$ and $f_2$ were chosen such that $f_m=|f_1 - f_2|$ caused the Raman vibrations to excite the first flexural resonance $f_{0,1}$ of the cantilever. We note that while the experimental setup in the inverted transmission geometry used a dichroic beamsplitter to combine the pump and stimulating beams, the experimental setup presented in Fig. 1 (b) does not require the use of a dichroic beamsplitter to combine the pump and stimulating beams. Indeed, we anticipate this technique might be useful in pump-probe experiments or multiple beam experiments in which it might be difficult to obtain the required dichroic optical components, particularly in experiments involving one or more infrared beams.

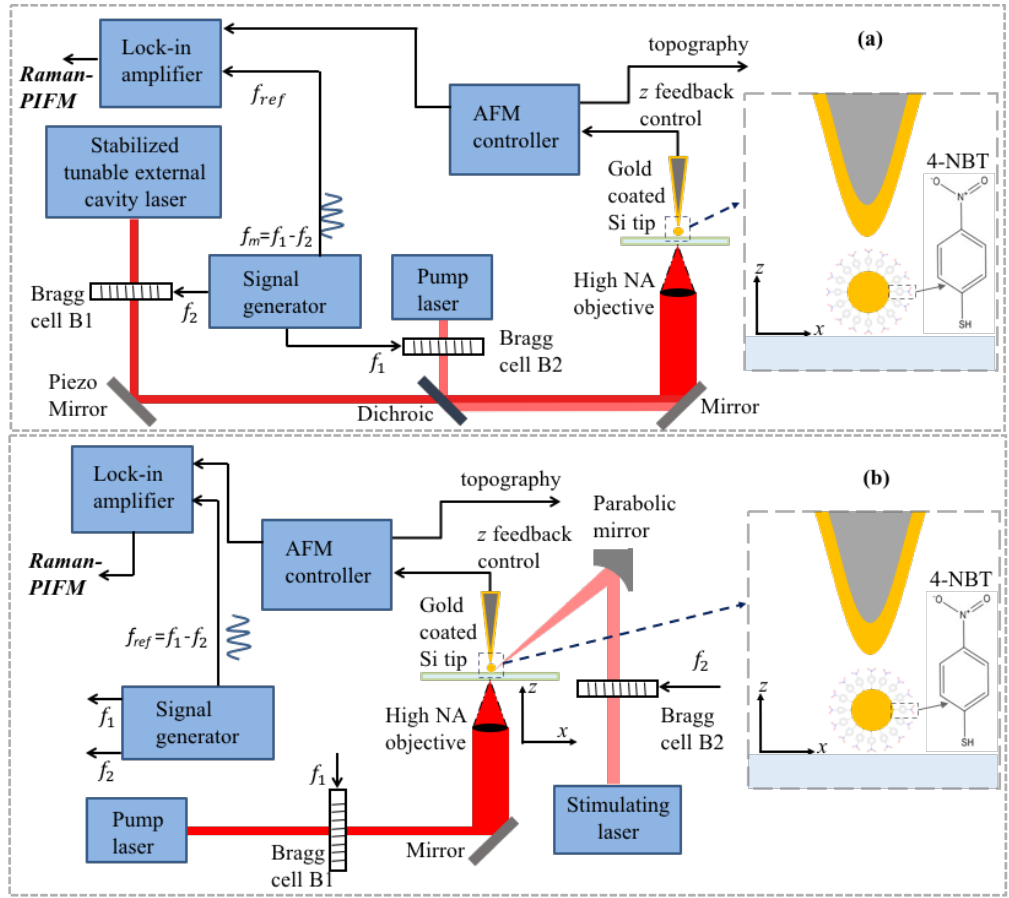

Fig. 1. Schematic of the experimental setups used to measure Raman-PIFM on gold nanoparticles functionalized with a SAM of 4-NBT molecules. (a) Setup with both the pump and the stimulating beam focused using a bottom objective in an inverted transmission geometry (b) Raman-PIFM configuration with the pump focused using a bottom objective in an inverted transmission geometry but with the stimulating beam focusing from the side at a 30° angle using a parabolic mirror.

Both configurations, detailed in Fig. 1, were built around an existing commercial AFM (VistaScope, Molecular Vista Inc.). In all experiments, Acousto Optic Modulators (AOM) were used to modulate the pump and the probe beams at frequencies $f_1$ and $f_2$, respectively. The modulation frequencies $f_1$ and $f_2$ were chosen such that each frequency individually did not coincide with any of the AFM cantilever's higher order eigenmodes. The pump laser was a helium-neon laser (wavelength 604 nm, Research Electro Optics) while the stimulating laser was a temperature stabilized 658 nm diode laser with the wavelength having a temperature tunable range from 656.5 nm to 664 nm. Commercial AFM cantilever probes (Nanosensors NCHR 300 kHz noncontact AFM cantilever) were prepared by sputter coating (South Bay Technology Ion Beam Sputtering/ Etching (IBS/e) System) of a 2 nm chromium adhesion layer followed by a 25 nm thick gold film. In all AFM experiments, we first drop cast a 10 µL drop of 1 nM aqueous solution of 4-NBT coated gold nanoparticles onto a UV-ozone cleaned glass coverslip with a thickness of 0.16 mm. Note that the same stock solution of 4-NBT coated gold nanoparticles was used to obtain results in both Figs. 3, 4 and 5.

**Results**

For the study, we functionalized gold nanoparticles (AuNPs) with a Raman reporter molecule (4-NBT). This Raman reporter molecule was selected not only for its strong Raman vibrational response at 1343 cm$^{-1}$ and but also because the line at 1343 cm$^{-1}$ is well separated from other Raman lines, allowing for de-tuning of the frequency $\Delta v = v_p - v_s$ far away from the 1343 cm$^{-1}$ resonance without exciting any other vibrational resonances of the molecule. The prepared functionalized AuNPs were characterized by measuring the Raman spectrum of the particles drop cast on a silicon substrate (Fig. 2 (a)). The Raman spectra of an aqueous solution of 4-NBT molecules drop cast on a silicon substrate and citrate-capped gold nanoparticles without 4-NBT drop cast on a silicon substrate are also shown in Fig. 2 (a). The Raman spectra indicate that the functionalization of AuNPs by 4-NBT molecules was successful and on average the sample clearly exhibited a plasmonic enhancement of Raman scattering from 4-NBT due to the gold nanoparticles. Fig. 2 (b) presents a Transmission Electron Microscope (TEM) image of the gold nanoparticles clearly showing the details of the faceted nature of the synthesized nanoparticles. From the TEM image we clearly identify single AuNPs in addition to clusters probably formed during sample preparation.

AuNPs were synthesized via the Fren's [24] method to yield particles in the range of 20-40 nm with a Localized Surface Plasmon Resonance (LSPR) peak at 532 nm (inset of Fig. 2 (b)). The synthesis method used was as follows: 291 µM solution of gold chloride trihydrate was heated to reflux, after which a 77.2 mM solution of sodium citrate tribasic dehydrate was added. The solution was allowed to reflux for about 20 mins to ensure that all sodium citrate was used. The resulting solution was a deep red/maroon color. To create a SAM of (4-NBT) molecules on the gold nanoparticle surface, a 15 µM solution of 4-nitrobenzene thiol was added to the gold nanoparticle solution and left at room temperature for 24 hours to ensure adequate functionalization. Next, the solutions of (4-NBT) and AuNPs were washed [25]-[28] by centrifugation at 7000 rpm for about 10 mins (Thermo Scientific Sorvall ST16R). The supernatant was removed and nanopure water (Barnstead Nanopure Diamond UV-UF, 18.1 MΩ/cm) was added to make 1 mL of gold nanoparticle/4-nitrobenzene thiol solution. All far-field Raman spectral measurements were made using a Process Instruments PI-200 Raman Micro spectrometer using a diode laser operating at 785 nm, at a power of 11 mW and with a 60 second acquisition time. The diameter of the focused beam spot was 832 μm.

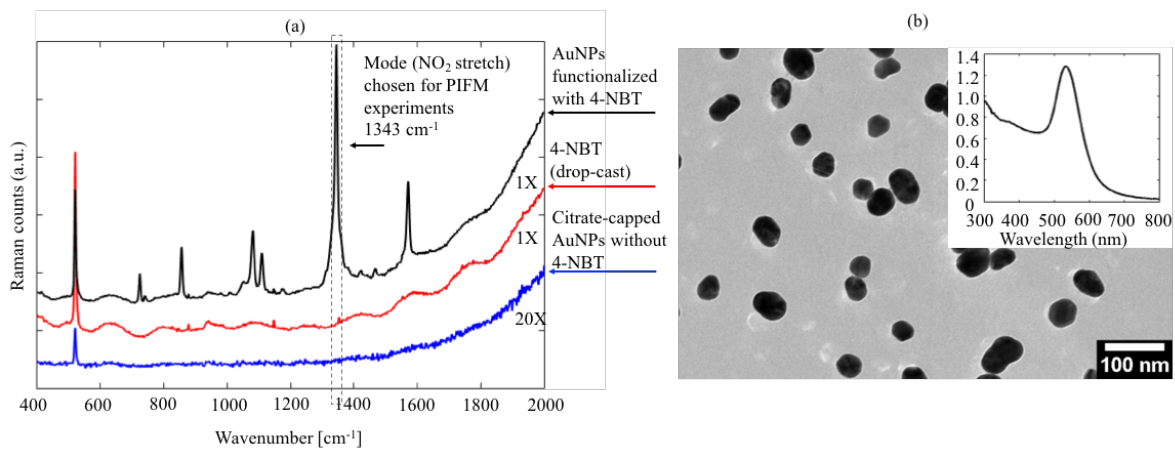

Fig. 2. (a) Raman spectra showing the signal enhancement provided by the gold nanoparticles in detecting the 4-NBT. The spectra are for (black) AuNPs functionalized with 4-NBT and drop-cast on a silicon substrate; (blue) Citrate-capped AuNPs drop-cast on silicon substrate without any 4-NBT; (red) 4-NBT molecules drop-cast on a silicon substrate without gold nanoparticles. The Raman peak at 1343 cm$^{-1}$ chosen for this experiment is highlighted. (b) TEM image of the gold nanoparticles showing the faceted nature of the nanoparticles. The polydispersity in size and shape of the gold nanoparticles contributes to the broadness of the localized surface plasmon resonance (LSPR) peak shown in the inset.

In Fig. 3 and 4, Raman-PIFM nanoscopy measurements performed using the inverted transmission geometry, detailed in Fig. 1 (a), are presented. The images represent experiments performed using two different gold coated AFM probes (300k NCHR), tip A and B, respectively, on AuNPs functionalized by a SAM of 4-NBT. Fig. 3 (a) and (b) show simultaneously recorded spatial distributions of topography and normalized Raman-PIFM images, respectively, when the wavelength of the stimulating laser was tuned such that $\Delta \nu = \nu_p - \nu_s$ coincided with the 1343 cm$^{-1}$ Raman vibrational mode. Fig. 3 (c) and (d) present the simultaneously recorded spatial distributions of topography and normalized Raman-PIFM images, respectively, when the wavelength of the stimulating laser was tuned such that $\Delta \nu = \nu_p - \nu_s$ was 1420 cm$^{-1}$. The lock-in time constant was 20 ms for this experiment. Line traces of the topography and normalized Raman-PIFM were extracted from Fig. 3 (a) (along line a-a') and Fig. 3 (b) (along line b-b') and plotted in Fig. 3 (e) and 3 (f), respectively. The traces plot the topography and Raman-PIFM of a single gold nanoparticle, of height 42 nm measured from topography plotted in Fig. 3 (a), functionalized with 4-NBT. The full width half maximum (FWHM) of the smallest feature in the line trace of topography is about 80 nm. The broadening in the line trace of topography is attributed to the convolution of the gold nanoparticle with the probe apex geometry. The FWHM of the feature in the Raman-PIFM is smaller at about 68 nm indicating an improvement in spatial resolution compared to topography. In the topography images plotted in Figs. 3 (a) and (c) we also observe a broad topological feature diagonally above the nanoparticle which could be nanoscale dust or residue left behind on the surface due to the sample preparation process.

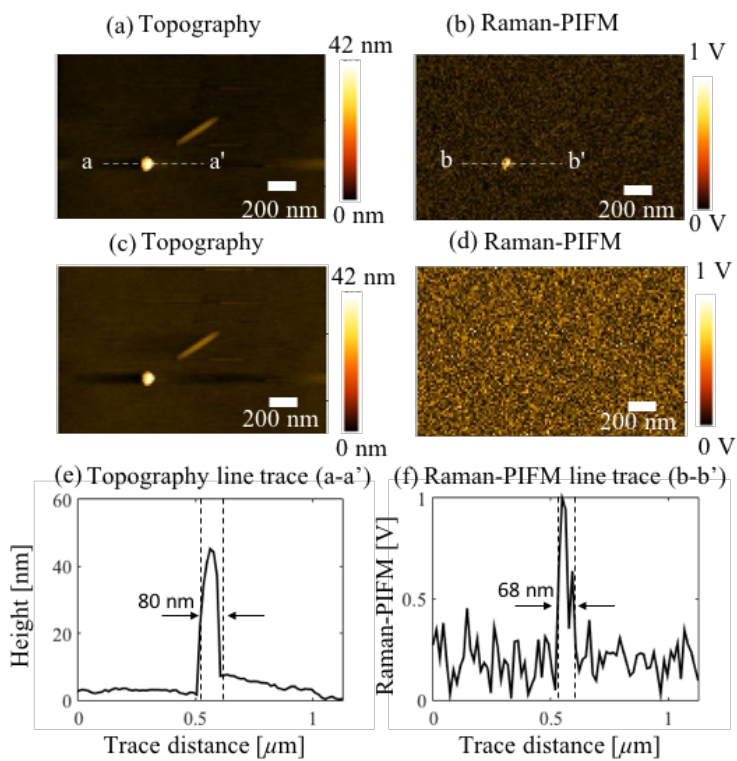

Fig. 3. Spatial distributions of (a) topography and (b) normalized Raman-PIFM spectral information of gold nanoparticle functionalized with a SAM of 4-NBT molecules measured simultaneously on resonance at 1343 cm$^{-1}$ Raman vibrational modes. Spatial distributions of (c) topography and (d) normalized Raman-PIFM spectral information of gold nanoparticle functionalized with a SAM of 4-NBT molecules measured simultaneously off resonance at 1420 cm$^{-1}$ (e) Line trace of topography (a-a') from Fig. 3 (a), (f) Line trace of Raman-PIFM (b-b') from Fig. 3 (b)

To verify the results in Fig. 3, we chose to repeat the Raman PIFM experiment on a different, freshly prepared sample with a different gold coated tip. The results of the Raman PIFM experiment are plotted in Fig. 4 (a) and (b), and show simultaneously recorded spatial distributions of topography and normalized Raman-PIFM images, respectively, when the wavelength of the stimulating laser was tuned such that $\Delta \nu = \nu_p - \nu_s$ coincided with the 1343 cm$^{-1}$ Raman vibrational mode. Fig. 4 (c) and (d) present the simultaneously recorded spatial distributions of topography and normalized Raman-PIFM images, respectively, when the wavelength of the stimulating laser was tuned such that $\Delta \nu = \nu_p - \nu_s$ was 1420 cm$^{-1}$. The lock-in time constant was 20 ms for this experiment.

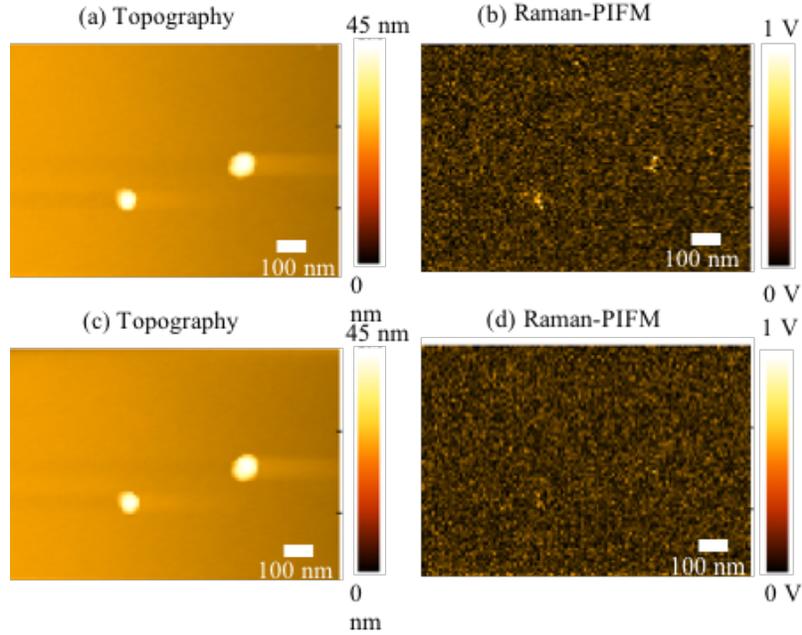

Fig. 4. Spatial distributions of (a) topography and (b) normalized Raman-PIFM spectral information of AnNPs functionalized with a SAM of 4-NBT molecules measured simultaneously on resonance at 1343 cm$^{-1}$ Raman vibrational modes. Spatial distributions of (c) topography and (d) normalized Raman-PIFM spectral information of the same NBT-functionalized AuNPs measured simultaneously off resonance at 1420 cm$^{-1}$.

To obtain results in both Figs. 3 and 4, the optical power of each beam (after modulation) just before the entrance pupil of the objective was adjusted to be 40 $\mu$W. To obtain results in Fig. 3, we chose an AFM cantilever with $f_{0,1} \sim 280$ kHz, $f_{0,2} \sim 1.8$ MHz. The modulation frequencies were chosen as $f_1 = 2.0$ MHz and $f_2 = 2.280$ MHz such that $f_m = f_2 - f_1 = f_{0,1}$. To obtain results in Fig. 4, we chose an AFM cantilever with $f_{0,1} \sim 260$ kHz, $f_{0,2} \sim 1.7$ MHz. The modulation frequencies were chosen as $f_1 = 2.0$ MHz and $f_2 = 2.260$ MHz such that $f_m = f_2 - f_1 = f_{0,1}$. To obtain results in Fig. 5, the optical power of the pump beam (after modulation) just before the entrance pupil of the objective was adjusted to be 40 $\mu$W. The optical power of the stimulating beam (after modulation) just before the parabolic mirror was adjusted to be 3.0 $m$W. We chose an AFM cantilever with $f_{0,1} \sim 280$ kHz, $f_{0,2} \sim 1.8$ MHz. The modulation frequencies were chosen as $f_1 = 2.0$ MHz and $f_2 = 2.280$ MHz such that $f_m = f_2 - f_1 = f_{0,1}$.

Having verified the Raman PIFM results in Figs. 3 and 4, we next investigated the alternate optical excitation geometry with non-collinear pump and probe beams described in Fig. 1(b). In Fig. 5, we present results obtained from Raman-PIFM nanoscopy experiments performed on NBT-functionalized AuNPs obtained using the combination of inverted transmission and the side illumination geometry detailed in Fig. 1 (b). Fig. 5 (a) and (b) show simultaneously recorded spatial distributions of topography and normalized Raman-PIFM images, respectively, when the wavelength of the stimulating laser was tuned such that $\Delta v = v_p - v_s$ coincided with the 1343 cm$^{-1}$ Raman vibrational mode. Fig. 5 (c) and (d) plot the simultaneously recorded spatial distributions of topography and normalized Raman-PIFM images, respectively, when the wavelength of the stimulating laser was tuned such that $\Delta v = v_p - v_s$ was 1420 cm$^{-1}$. The lock-in time constant was 20 ms for this experiment.

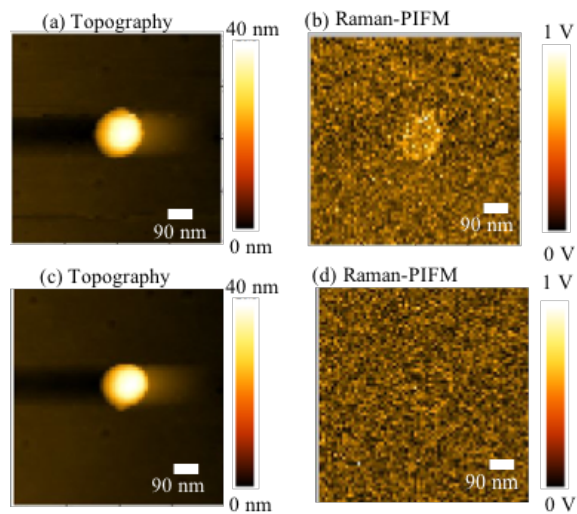

Fig. 5. Spatial distributions of (a) topography and (b) normalized Raman-PIFM spectral information of an AuNP functionalized with a SAM of 4-NBT molecules measured simultaneously on resonance at 1343 cm$^{-1}$ Raman vibrational modes. Spatial distributions of (c) topography and (d) normalized Raman-PIFM spectral information of the NBT-functionalized AuNP measured simultaneously off resonance at 1420 cm$^{-1}$.

In Figs. 3(b), 4(b) and 5(b), we clearly observe the Raman vibrational responses from the 4-NBT on the AuNPs while on vibrational resonance while, in Figs. 3(d), 4(d) and 5(d), the Raman-PIFM signal was completely diminished to the noise level for off vibrational resonance excitation of the molecules confirming the measurement of Raman vibrational responses of molecules using photon-induced forces.

**Discussion**

We first briefly comment on the importance of tip preparation for Raman-PIFM measurements. Comparison of Raman-PIFM images plotted in Fig. 3 (b), 4 (b) and 5 (b) shows better signal-to-noise ratio for the photon induced force data in Fig. 3 (b) compared to the data plotted in Figs. 4 (b) and 5 (b). We note that the reasons for this change could be due to: (1) variation of quality among different gold-coated tips used for the Raman-PIFM experiment and/or (2) the modification of tips during scans and/or (3) a variation in the number of thiol molecules attached to different nanoparticles. We believe the variation of quality, measured as PIFM response, among different tips is due to the sputtering process used for gold coating of AFM cantilever probes. The tips were fabricated by sputtering gold onto a commercial silicon cantilever. During the sputtering process, the tip was mounted on a stage which was both periodically rotated and tilted at a slow pace. However, sputtering processes causes the growth of large grains of gold which, sometimes, may not be well defined at the apex of the silicon cantilever. This ambiguity regarding the size and position of the gold grain at the apex of the tip causes variation in the PIFM response of sputter-coated tips.

We minimized the variation in quality among tips by characterizing each tip before use in the Raman-PIFM experiment. Care was taken to ensure that each tip was tested using the same conditions and the tip was not damaged during testing. We used the PIFM technique to test each tip. We applied the image force technique outlined in [11] by using the tip to map the focal fields of a linearly polarized, tightly focused optical beam to obtain double lobes. The following metric was used to compare different tips: we computed the ratio of the peak PIFM signal value of the double lobe (in $\mu$V) to the average noise in the PIFM signal with no optical excitation (in $\mu$V). For all tests, the optical power of the pump beam (after modulation) just before the entrance pupil of the objective was 100 µW and the wavelength of light for all tests was 604 nm. Only tips whose measured ratios were 50 or greater were selected for further use in Raman-PIFM experiments. Indeed, the results plotted in Figs. 3, 4 and 5 were obtained with tips with ratios of 100, 70 and 50, respectively. The results plotted in Figs. 3, 4 and 5 clearly establish the importance of the preparation of the gold coated AFM probe for the Raman-PIFM experiment. Therefore, it would be very useful to fabricate tips with precise, well-defined and highly reproducible gold nano-structure at the apex of the cantilever. One possibility could be to attach spherical metal nanoparticles to the end of a silicon cantilever [17], [18] thereby eliminating any ambiguities caused by sputter coating of tips.

To estimate the number of molecules detected by the Raman-PIFM experiment, we first calculated the number of NBT molecules expected to be bound to a AuNP assuming formation of a SAM and used this to estimate the number of molecules excited by the tip-enhanced stimulated Raman process. Indeed, the maximum bound on the number of detected molecules is equal to or lesser than the number of molecules excited. To compute the number of molecules excited by the tip-enhanced stimulated Raman process, denoted as $N_M$, we computed the area density of 4-NBT molecules on gold, denoted $\rho_M$, and the area on a plane 0.5 nm above the gold nanoparticle excited by the third order, non-linear, Stimulated Raman Scattering (SRS) process, denoted as $A_E$. To compute $A_E$, we ran three dimensional electromagnetic simulations using the commercial finite-element code COMSOL Multiphysics at the pump and stimulating wavelengths of 604 nm and 657 nm, respectively, and detailed in the Appendix.

We note that the total time averaged, $z$ directed, optical force is given by $\langle F_z \rangle \propto Re\left[\int d\mathbf{r} \mathbf{P}^{(3)}(\mathbf{r}) \cdot \nabla E_{z,STIM}^*(\mathbf{r})\right]$ [9]-[11], where, $P^{(3)}$ is the induced third order non-linear polarizability $P^{(3)} \propto |E|^3$ which is excited by the two beams: pump and probe [19]. In Fig. 6 (a), we plot the maximum value of electric field enhancement in the tip-sample gap, normalized to field intensity of incident exiting wave, calculated from COMSOL simulations (described in the Appendix). The enhancement in the electric field in the tip-sample gap was extracted from a plane 0.5 nm above the gold nanoparticle at the pump and stimulating wavelengths. Considering $P^{(3)} \propto |E_{z,PUMP}||E_{z,STIM}|^2$, we plot the maximum value of the polarization $P_{max}^{(3)}$ as a function of the tip sample gap in Fig. 6 (b). Using the procedure outlined in the Appendix, the number of molecules excited by the tip-enhanced stimulated Raman process as a function of tip-sample gap is plotted in Fig. 6 (c).

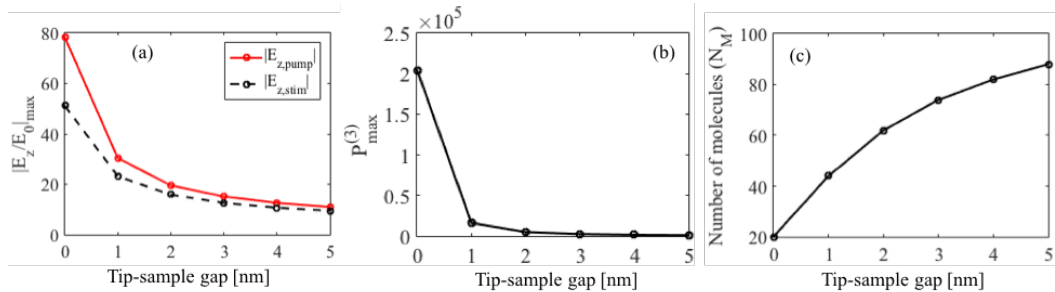

Fig. 6. (a) Calculated maximum electric field enhancement between the tip and nanoparticle as a function of tip-sample gap, normalized to field intensity of incident excited wave, extracted from a plane 0.5 nm above the gold nanoparticle at the pump and stimulating wavelengths (b) Calculated maximum value of polarization $P_{max}^{(3)}$ using $P^{(3)} \propto |E_{z,PUMP}||E_{z,STIM}|^2$ as a function of tip-sample gap and (c) Number of molecules excited by the overlapping pump and stimulating beams, $N_M = A_E \rho_M$, where the area was computed from COMSOL simulations using procedure outlined in Appendix.

From Fig. 6 (b), we note the polarization ratios $P^{(3)}_{tsgap=0\,nm}/P^{(3)}_{tsgap=1\,nm} \sim 12.5$, $P^{(3)}_{tsgap=0\,nm}/P^{(3)}_{tsgap=2\,nm} \sim 41.5$ and $P^{(3)}_{tsgap=0\,nm}/P^{(3)}_{tsgap=3\,nm} \sim 85.5$ (here, tsgap means tip-sample gap) indicating, as expected, that reducing the gap increased the excitation electric field strength and thereby producing a corresponding increase in detected signal. Therefore, due to the intermittent contact mode of AFM operation used in this experiment and high field enhancement and subsequently large value of the ratios $P^{(3)}_{tsgap=0\,nm}/P^{(3)}_{tsgap=j\,nm}$, where, $j = 1, 2, 3$, etc., for tip-sample gap = 0 nm, we expect only those molecules within the area given by $A_{E,gap=1\,nm} \sim 10$ nm$^2$, when the tip-sample gap = 0 nm, to dominate the measured Raman-PIFM signal. Therefore, we use the area $A_{E,gap=1\,nm} \sim 10$ nm$^2$ to calculate the number of molecules excited by the tip-enhanced stimulated Raman process.

Using the packing density of the 4-NBT molecules (2 molecules/nm$^2$ [20], [21]), we calculate the total number of 4-NBT molecules packed onto the surface of a single AuNP. We find this value to be about 10,000 molecules. We note that the estimated packing density of 4-NBT molecules on a gold surface used in our calculations assumes the 4-NBT molecules are packing on an atomically flat gold surface thereby affecting our calculated value. The actual packing density could be smaller for the AuNPs due to the curvature, different facets and defect sites. Based on these assumptions, an estimate of the maximum bound on the number of molecules excited by the tip-enhanced stimulated Raman process is $N_M \sim 20$ molecules for tip-sample gap = 0 nm. Since, the maximum bound of number of detected molecules is equal to or lesser than the maximum bound on the number of excited molecules, we estimate that the maximum number of molecules detected by the tip at the position of maximum field enhancement is about 20 molecules.

After performing Raman-PIFM experiments on many nanoparticles from the same batch with the same tip we found that about 30% of the functionalized gold nanoparticles did not give any Raman-PIFM signal. This could result from a smaller than expected number of 4-NBT molecules on the surface of the nanoparticle due to incomplete 4-NBT monolayer and an over estimation of the 4-NBT packing density [22], [23]. Indeed, it has been shown that not all citrate molecules are easily displaced from the surface of AuNPs during functionalization, even with thiols [22], [23]. We also note, in the simulations we considered smooth spheres without facets. However, as evident from TEM image in Fig. 2(b), the AuNPs are faceted features with sharp edges which can cause higher field enhancements than predicted by simulations. This implies that the signal arising from molecules distributed over a small area could, in experiment, arise from few molecules near sharp edges on the gold nanoparticles. Therefore, the estimated maximum number of molecules detected by the tip at the position of maximum field enhancement could be fewer than 20 molecules.

For a tip-sample gap = 1 nm, we note from Fig. 6 (b), the polarization ratio $P^{(3)}_{tsgap=0\,nm}/P^{(3)}_{tsgap=1\,nm} \sim 12.5$ and $E_{z,STIM,tsgap=0\,nm}/E_{z,STIM,tsgap=1\,nm} \sim 2.5$. This indicates the measured time averaged force signal $\langle F_z \rangle$ is reduced by $\sim 31$ (using $\langle F_z \rangle \propto Re[\int d\boldsymbol{r} \boldsymbol{P}^{(3)}(\boldsymbol{r}) \cdot \nabla \boldsymbol{E}^*_{z,STIM}(\boldsymbol{r})]$) when the tip-sample gap was changed from 0 nm to 1 nm. However, the number of molecules excited by the tip-enhanced stimulated Raman process, $N_M$, only doubled from $\sim 20$ molecules for tip-sample gap of 0 nm to $\sim 40$ molecules for tip-sample gap of 1 nm. This result suggests that few molecule resolution in Raman-PIFM, allowing for study of local chemistry at nanoscale interfaces, can be achieved even with a tip-sample gap as large as 1 nm provided the detection scheme is sensitive enough to compensate for reduced signal level (when compared to operating at a tip-sample gap of 0 nm) due to the increased tip-sample gap of 1 nm. However, we note that force detection scheme can be further improved by (a) placing the experimental setup in a rough vacuum and thereby increasing the *Q* of the cantilever 100 times and (b) lowering the operating temperature from 300 K to 3 K, the minimum detectable force can be greatly reduced by several orders of magnitude from the current pico-newton range to atto-newton range [14]. Performing Raman-PIFM on interfaces with a tip-sample gap as large as 1 nm could allow for sensitive Raman-PIFM measurements to be performed while not allowing the tip to be contaminated/damaged by contact with the sample surface thereby potentially reducing the modification of tips during a scan.

In conclusion, we have demonstrated the stimulated Raman nanoscopy of a few molecules excited and measured only using near-field photon induced forces without the need for far-field optical spectroscopic detection. Furthermore, molecules were excited without the need for any resonant electronic enhancement gain. The molecules were imaged by vibrating an AFM cantilever on one of its second flexural eigenmodes enabling the tip to be controlled much closer to the sample thereby improving the detection sensitivity. We imaged ~30 nm diameter AuNPs functionalized with a SAM of 4-NBT molecules. Using simulations, we estimated the maximum number of molecules detected by the tip at a position of maximum field enhancement to be about 20 molecules. We also demonstrated a novel implementation of Raman-PIFM nanoscopy with non-collinear pump and stimulating beams. The results of this work were obtained in ambient conditions, using only continuous-wave lasers and without the use of ultrafast pulsed lasers. The use of ultrafast pulsed lasers and a rough vacuum to operate the AFM could push the detection sensitivity towards the single molecule limit. The ability to detect the Raman responses of a few molecules using only AFM mechanical detection instead of photon collection could allow for sensitive means to study local chemistry at interfaces with nanoscale resolution.

**Appendix**

All electromagnetic simulations were performed using the RF module of the commercial Finite Element Method solver COMSOL Multiphysics. The gold coated AFM probe was approximated by a gold nanoparticle of diameter 50 nm while the diameter of the sample gold nanoparticle was taken to be 28 nm, which was the diameter of the smallest measured nanoparticle by Raman-PIFM (see Fig. 4 (a)). The excitation of the tip-sample junction was done by interfering two obliquely incident plane waves at the junction, with the plane waves traveling in the +z direction but traveling from the substrate (refer to Fig. 6 (a) for definitions). The simulation geometry and approach was very similar to our approach followed in [14]. The optical constants of gold used in all calculations in the work were obtained from [29]. As a first order approximation, we ignore any variation of gap electric field strength along the length of the molecule. The area was calculated on a plane that was placed 0.5 nm from the surface of the gold nanoparticle, which is about half the length of the molecule (4-NBT).

To compute $A_E$, we ran three dimensional electromagnetic simulations using the commercial finite-element code COMSOL Multiphysics at the pump and stimulating wavelengths of 604 nm and 657 nm, respectively, for different gaps sizes from 1 nm to 5 nm while keeping all other simulation parameters constant. The gold coated AFM probe-AuNP gap was represented by the gap between two AuNPs to simplify the model for the simulations. Note that a gap size of 1 nm in simulation corresponded to the AFM probe tip just touching the 4-NBT molecule on the AuNP and we expect this case to have the highest field enhancement. The gap was set at 1 nm which is roughly the length of 4-NBT molecule and this corresponds to a tip-sample gap of 0 nm in an actual experiment. A two-dimensional cross section plot of the unity normalized, absolute value of electric field distributions in the gap ($|E_z|$) computed by COMSOL for the case of 1 nm gap is plotted in Fig. 7 (a). For the geometry of excitation used in this work, we expect the maximum field enhancement to occur at the spatial positions of the particles as shown in Fig. 7 (a) as confirmed by simulations. From the simulations, the spatial distribution of $|E_z|$ on a plane 0.5 nm above the gold nanoparticle at the pump and stimulating wavelengths, $|E_{z,PUMP}|$ and $|E_{z,STIM}|$, respectively, were extracted.

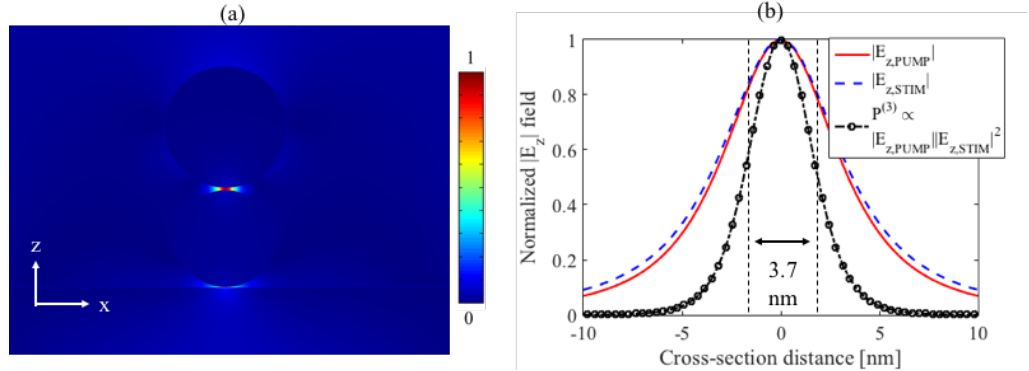

Fig. 7. (a) Spatial distributions of unity normalized intensity distribution between two gold nanoparticles of diameters 50 nm and 28 nm separated by a gap of 1 nm. (b) Plot of electric fields of the pump and stimulating beams extracted from COMSOL simulations. Also plotted is the product $|E_{z,PUMP}||E_{z,STIM}|^2$ which gives diameter of hotspot ~ 3.7 nm.

The excitation area, $A_E$, can be estimated by considering the induced third order non-linear polarizability $P^{(3)} \propto |E|^3$ which is excited by the two beams: pump and probe [19]. Note that $P^{(3)}$ will have terms from various possible combinations of fields such as $|E_{z,PUMP}|^2|E_{z,STIM}|$ and $|E_{z,PUMP}||E_{z,STIM}|^2$. Considering $P^{(3)} \propto |E_{z,PUMP}||E_{z,STIM}|^2$, we plot the normalized term $|E_{z,PUMP}||E_{z,STIM}|^2$ for a gap size of 1 nm in Fig. 7 (b) and find the Full Width Half Maximum (FWHM), or the diameter of the plasmonic field enhancement region or "hotspot" between the two gold nanoparticles to be ~ 3.7 nm giving an excitation area, $A_{E,gap=1\ nm} \sim 10\ nm^2$.

**Acknowledgements**

This work was supported by the NSF Center for Chemical Innovation, Chemistry at the Space-Time Limit (CaSTL) under Grant No. CHE-1414466. The authors thank Prof. V. Ara Apkarian, Prof. Eric O. Potma and Dr. Sudipta Mukherjee from the Department of Chemistry, University of California, Irvine, Dr. Fei Huang from Department of Electrical Engineering and Computer Science, University of California, Irvine, Dr. Derek Nowak, Will Morrison and Dr. Sung Park from Molecular Vista Inc. for helpful discussions. We acknowledge Dr. Jian-Guo Zheng in in the UC Irvine Materials Research Institute (IMRI) for helpful discussion. We acknowledge South Bay Technology for allowing us to use its Ion Beam Sputtering/Etching (IBS/e) System) in the preparation of AFM cantilever probes in the UC Irvine Materials Research Institute (IMRI)